\input{aipcheck}

\documentclass[
    ,final            % use final for the camera ready runs
  ]
  {aipproc}

\layoutstyle{6x9}

\renewcommand\XFMtitleblock{%
  \XFMtitle
  \let\XFMoldpar\par
  \def\par{\XFMoldpar\def\par{\space
             (for the CTA Consortium)\XFMoldpar}}%
   \XFMauthors
   \let\par\XFMoldpar
   \XFMaddresses
   \XFMabstract
   \vspace{5pt}%
   \XFMkeywords
   \XFMclassification
 }
\usepackage{amsmath}
\usepackage{url}

\begin{document}

\title{Monte Carlo comparison of mid-size telescope designs for the Cherenkov Telescope Array}


\author{T. Jogler}{
  address={SLAC National Accelerator Laboratory, 2575 Sand Hill Road M/S 29,
Menlo Park, CA 94025, USA}
}

\author{M. D. Wood}{
  address={SLAC National Accelerator Laboratory, 2575 Sand Hill Road M/S 29,
Menlo Park, CA 94025, USA}
}

\author{J. Dumm}{
  address={University of Minnesota, 116 Church St SE, Minneapolis, MN 55455, USA}
}

\begin{abstract}
  The Cherenkov Telescope Array (CTA) is a future very high energy
  gamma-ray observatory. CTA will be comprised of small-, medium- and
  large-size telescopes covering an energy range from tens of GeV to
  hundreds of TeV and will surpass existing telescopes in sensitivity
  by an order of magnitude. The aim of our study is to find the
  optimal design for the medium-size telescopes (MSTs), which will
  determine the sensitivity in the key energy range between a few
  hundred GeV to about ten TeV.  To study the effect of the telescope
  design parameters on the array performance, we simulated arrays of
  61 MSTs with 120~m spacing and a variety of telescope
  configurations.  We investigated the influence of the primary
  telescope characteristics including optical resolution, pixel size,
  and light collection area on the total array performance with a
  particular emphasis on telescope configurations with imaging
  performance similar to the proposed Davis-Cotton (DC) and
  Schwarzschild-Couder (SC) MST designs.  We compare the performance
  of these telescope designs, especially the achieved gamma-ray
  angular resolution and differential point-source sensitivity.

\end{abstract}

\maketitle

%%%%%%%%%%%%%%%%%%%%%%%%%%%%%%%%%%%%%%%%%%%%
%% MAINMATTER
%%%%%%%%%%%%%%%%%%%%%%%%%%%%%%%%%%%%%%%%%%%%

\section{Introduction}
The Cherenkov Telescope Array (CTA) is the future next generation
Imaging Atmospheric Cherenkov Telescope (IACT) observatory. CTA aims
to surpass the current IACT systems like HESS, MAGIC and VERITAS by an
order of magnitude in sensitivity and  enlarge the observable energy
range from a few tens of GeV to far beyond one hundred
TeV~\cite{whitepaper}. To achieve this broad energy range and high
sensitivity, CTA will be comprised of three different telescope
sizes. These are denoted according to their mirror diameter into
large-size telescopes, medium-size telescopes, and small-size
telescopes.

In this paper we investigate the effect of the optical point-spread
function and the camera pixel size on the achievable point-source
sensitivity. We investigate medium-size telescopes since they are most
sensitive in the energy range where the best angular resolution is
achieved and small pixels sizes are most feasible.

The main motivation for this study is to determine if the current,
well-tested single-mirror design (Davis-Cotton, DC) or a new two
mirror design (Schwarzschild-Couder, SC) would be the best choice for
the medium-size CTA telescopes.  The SC telescopes can achieve much
smaller optical point spread functions (PSF) but require many more
read out channels and more complicated mirror designs which increase
their price compared to a DC telescope with similar mirror area. We
simulate idealized telescope parameters for both designs and compare
their gamma-ray PSF and point-source sensitivity.
  
\section{Simulations}

Gamma-ray and proton air showers were simulated with the CORSIKA Monte
Carlo (MC) package \cite{heck1998} and the QGSJet-II hadronic
interaction model \cite{2006NuPhS.151..143O}.  Simulations were
performed for an array at an elevation of 2000~m and geomagnetic field
configuration similar to the proposed southern hemisphere sites.
Showers were simulated at 20$^\circ$ zenith angle over the energy
range from 10~GeV to 30~TeV.  All simulations use the same array
layout comprising 61 telescopes forming a square with 120~m
inter-telescope spacing. Each array is composed of identical
telescopes.

\subsection{Telescope Designs}
We simulated a range of optical PSFs and pixel sizes that bracket the
imaging performance of the DC- and SC-like telescope designs.  The
proposed designs for the DC- and SC-MST have a 68\% optical PSF
containment radius (R$_{68}$) of 0.02$^\circ$-0.04$^\circ$ and
0.04$^\circ$-0.1$^\circ$ over the FoV and a pixel size
(R$_\text{pix}$) of 0.06$^\circ$ and 0.16$^\circ$.  We use the
configurations with R$_{68}$/R$_\text{pix}$ of
0.02$^\circ$/0.06$^\circ$ and 0.08$^\circ$/0.16$^\circ$ as
representative of configurations with DC- and SC-like imaging
performance respectively.  For both configurations we assume a field
of view of $8^{\circ}$.

After imaging resolution, the other important characteristic of the
telescope optical system is the total effective light collection area
($A_\text{opt}(\lambda)$), defined as the product of the mirror area
and the optical efficiencies of all telescope components in the
optical path including the mirrors and photosensors.  We simulated a
baseline telescope configuration with an aperture of 10~m and a mirror
area of 78.5~m$^{2}$.  We assume a photosensor with a peak PDE of 24\%
at 380~nm and a spectral response similar to the Hamamatsu R1398.  
The baseline configuration has
$\left<A_\text{opt}\right> = 11$ m$^{2}$ between 250 nm and 700 nm for
a Cherenkov-like spectral distribution attenuated for an emission
height of 10~km.  To perform a realistic comparison of the DC and SC
designs at fixed cost we consider an SC configuration with
$A_\text{opt}$ reduced by a factor of 0.562.  This factor includes the
effect of both a smaller mirror area and a slightly higher photosensor
PDE which could be achieved with silicon photo-multipliers.

\subsection{Trigger and DAQ}
The trigger and readout electronics are not simulated in detail but
modeled such that the magnitude of their effect on the array
performance can be estimated.  The noise in each channel is modeled as
the sum of a Poisson-distributed NSB term and a Gaussian-distributed
electronics noise term with an RMS of 0.1~phe per channel.  The NSB
amplitude in each pixel is equal to $\Delta\Omega
\epsilon\sigma_{\text{NSB}}$ where $\Delta\Omega$ is the pixel solid
angle, $\epsilon = (A_\text{opt}/\tilde{A}_\text{opt})$ is the
effective light collection area relative to the baseline
configuration, and $\sigma_{\text{NSB}} = 100$~phe~deg$^{-2}$ is the
NSB density in the focal plane.  The baseline NSB density was chosen
to be representative for an extragalactic observation field and an
integration gate of 10~ns.  Each telescope camera containing more than
60~phe is assumed to trigger, and at least two telescopes must trigger
to produce an array trigger. All array triggered events are further
processed.

\subsection{Analysis}
Reconstruction of the telescope image data into event-level parameters
proceeds in three stages.  First, an image cleaning is performed to
select pixels with statistically significant signal amplitude.  The
shower trajectory is then reconstructed using a geometric analysis of
the moments of the light distribution in each camera.  Finally a
likelihood-based reconstruction is performed using templates for the
light distribution in each telescope derived from MC simulations.  In
addition to the event trajectory and energy, a number of parameters
useful for gamma-hadron discrimination are calculated such as the
goodness-of-fit of the telescope images with respect to the image
templates.  Background suppression is performed with the TMVA boosted
decision tree (BDT) method \cite{tmva}. 
Energy-dependent cuts on the BDT
output variable and $\theta^2$, the squared angular separation between
the reconstructed and source directions, are optimized under the
assumption of a point-like source distribution with an intensity equal
to 1\% of the Crab Nebula flux.

\section{Results \& Conclusions}

\begin{figure}[h]
  \includegraphics[width=.49\textwidth]{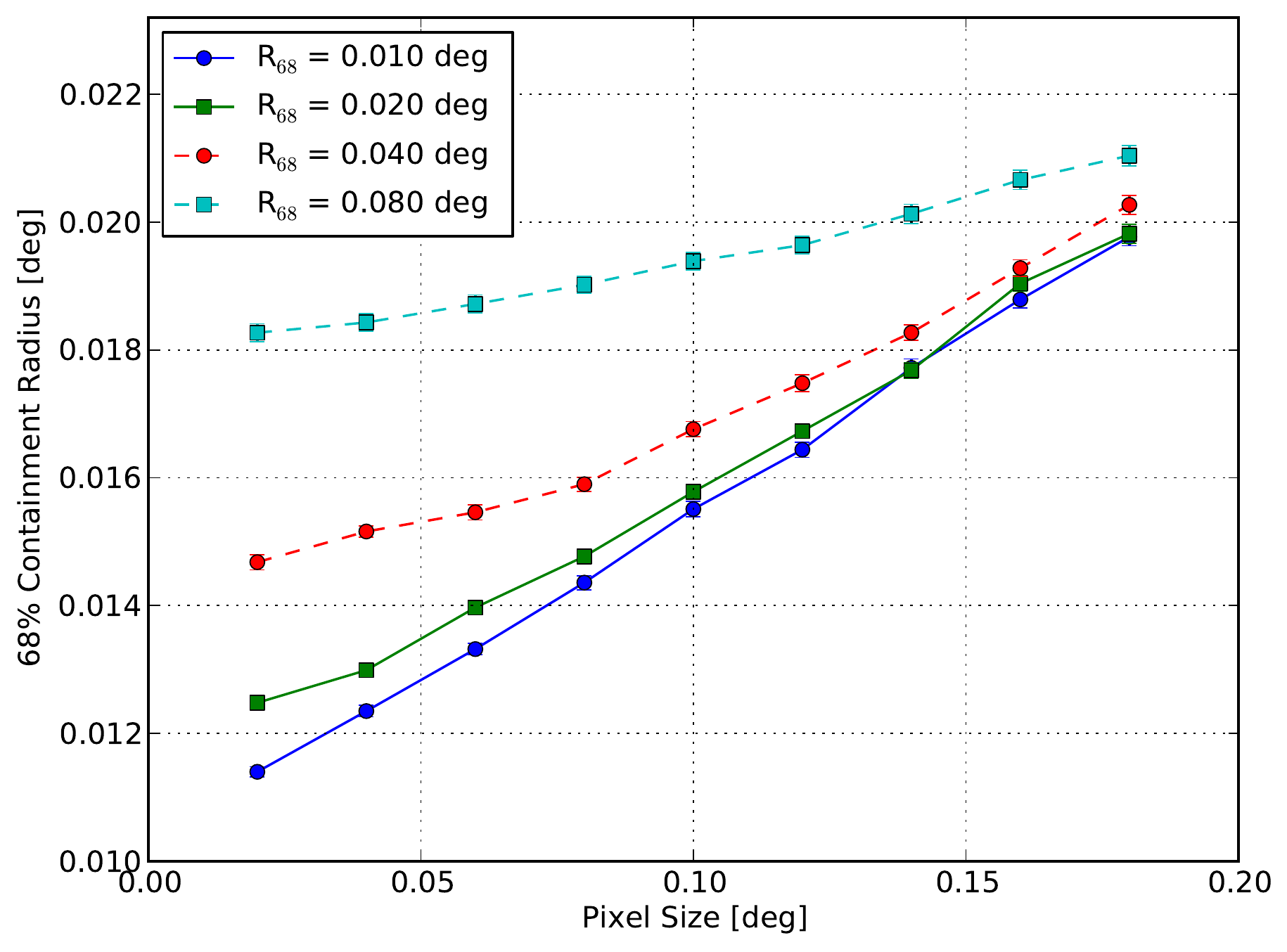}
  \includegraphics[width=.49\textwidth]{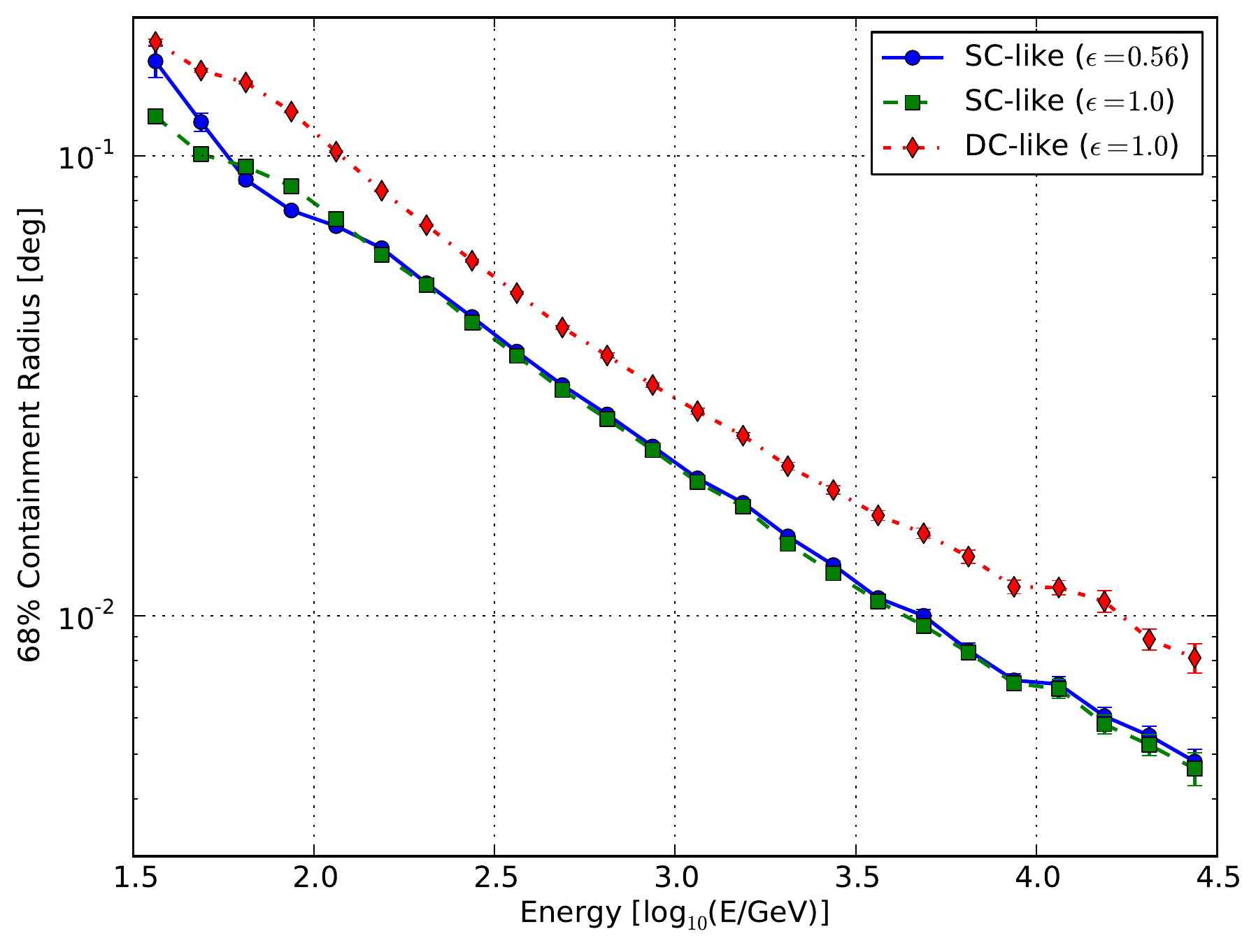}
  \caption{\textbf{Left:} 68\% containment radius of the gamma-ray PSF
    at 1~TeV versus pixel size shown for a 61 telescope MST array
    composed of telescopes with increasing 68\% optical PSF
    containment radii: 0.01$^\circ$ (blue circles and solid line),
    0.02$^\circ$ (green squares with solid line), 0.04$^\circ$ (red
    circles and dashed line), 0.08$^\circ$ (cyan squares and dashed
    line).  \textbf{Right:} 68\% containment radius of the gamma-ray
    PSF versus gamma-ray energy for an array composed of telescopes
    with SC-like imaging performance with effective light collection
    area scaled by 0.56 (blue circles and solid line) and 1.0 (green
    squares and dashed line) and a DC-like imaging performance (red
    diamonds and dot-dashed line).}

\end{figure}

\begin{figure}[h]
  \includegraphics[width=.49\textwidth]{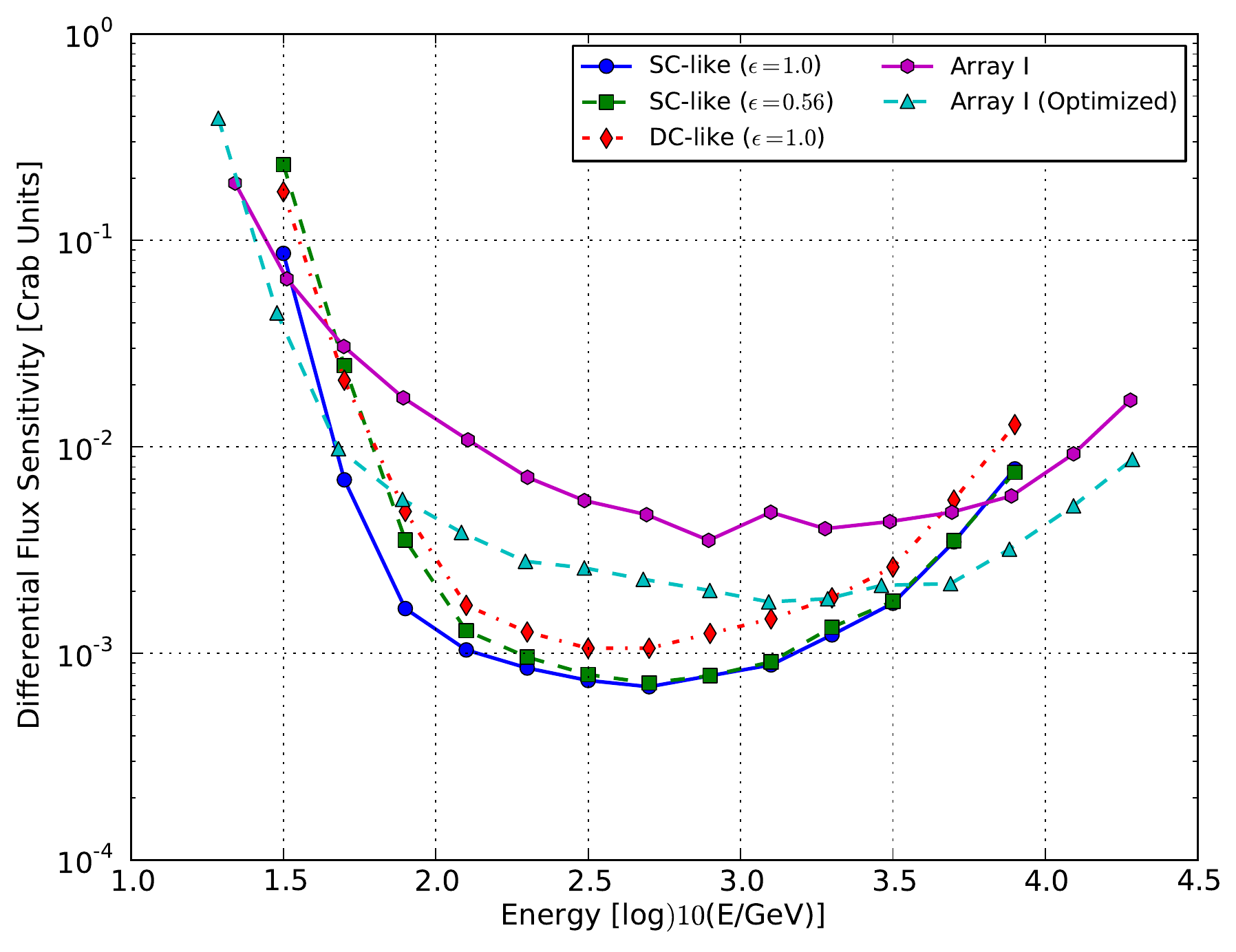}
  \includegraphics[width=.49\textwidth]{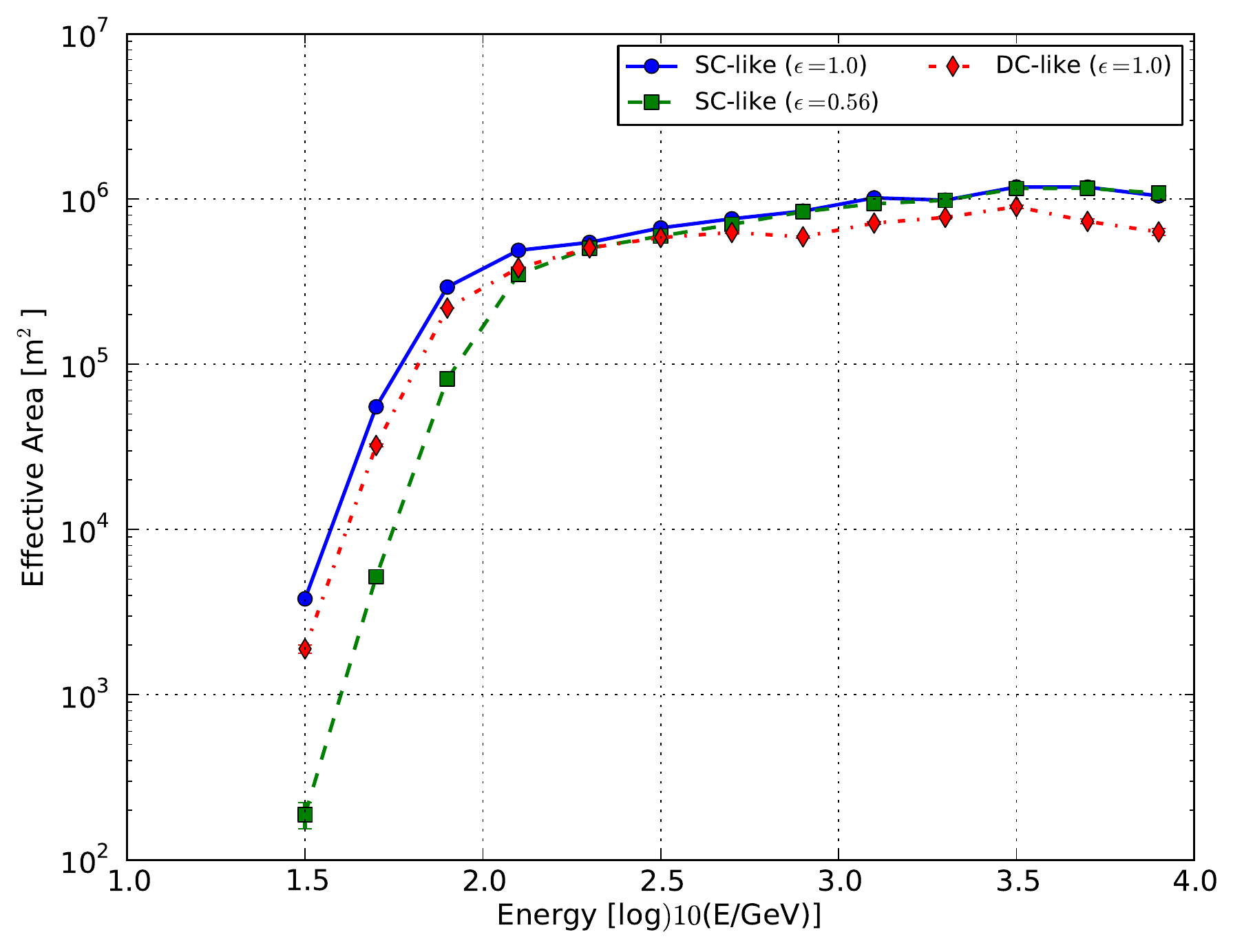}
  \caption{\textbf{Left:} Differential point-source sensitivity for a
    61 telescope array with DC- (red diamonds and dot-dashed line) and
    SC-like (green squares and dashed line) telescope designs with
    assumed same cost and one of the proposed CTA designs (array I)
    with 18 DC-like MSTs \cite{Bernlohr2012} for the baseline (magenta
    hexagons and solid line) and optimized (cyan triangles and
    dashed line) analyses.  The blue circles with solid line shows the
    performance of the SC-like telescope design when the light
    collection area is equal to that of the DC-like
    design. \textbf{Right:} Effective area versus gamma-ray energy for
    the three configurations shown in the left figure: DC-like (red
    diamonds and dot-dashed line), SC-like with $\epsilon = 0.56$
    (green squares and dashed line), and SC-like with $\epsilon = 1.0$
    (blue circles and solid line).}
\end{figure}

The gamma-ray PSF improves when reducing the pixel size as long as the
optical PSF is smaller than the pixel size. The SC-like telescope
array shows a 40\% improved gamma-ray PSF compared to the DC-like
telescopes at all energies.  As shown by the comparison of SC-like
arrays with $\epsilon = 1.0$ and $\epsilon = 0.56$ in Fig. 1, the
gamma-ray angular resolution is nearly independent of the telescope
effective light collection area except at very low energies.  Fig. 2
shows that the SC-like array has a $\sim$50\% better differential
sensitivity relative to the DC-like array at energies above 100~GeV
which is mainly due to the improved gamma-ray PSF.  Below 100~GeV the
smaller light collection area of the SC-like telescope configuration
is a disadvantage resulting in a higher reconstruction energy
threshold and an equal or slightly worse differential sensitivity.
While the SC-like array is more sensitive compared to the DC-like
array no SC telescope has been built to date. The results presented
here provide encouragement to build an SC prototype telescope to test
if the performance can be achieved under realistic conditions.

%%%%%%%%%%%%%%%%%%%%%%%%%%%%%%%%%%%%%%%%%%%%%%%%
%% BACKMATTER
%%%%%%%%%%%%%%%%%%%%%%%%%%%%%%%%%%%%%%%%%%%%%%%%

\begin{theacknowledgments}
We gratefully acknowledge support from the agencies and organizations 
listed in this page: \url{http://www.cta-observatory.org/?q=node/22}
\end{theacknowledgments}



\bibliographystyle{aipproc}   % if natbib is available

\bibliography{sample}

\IfFileExists{\jobname.bbl}{}
 {\typeout{}
  \typeout{******************************************}
  \typeout{** Please run "bibtex \jobname" to optain}
  \typeout{** the bibliography and then re-run LaTeX}
  \typeout{** twice to fix the references!}
  \typeout{******************************************}
  \typeout{}
 }

\end{document}